# Establishment of global phase coherence in a highly disordered fractal MgO/MgB$_2$ nanocomposite: Roles of interface, morphology and defect

Iku Nakaaki,[1] Aoi Hashimoto,[1] Shun Kondo,[2] Yuichi Ikuhara,[2,3] Shuuichi Ooi,[4] Minoru Tachiki,[4] Shunichi Arisawa,[5] Akiko Nakamura,[6] Taku Moronaga,[6] Jun Chen,[7] Hiroyo Segawa,[7] Takahiro Sakurai,[8] Hitoshi Ohta,[9] and Takashi Uchino[1,*]

[1] *Department of Chemistry, Graduate School of Science, Kobe University, Kobe 657-8501, Japan*
[2] *Institute of Engineering Innovation, School of Engineering, The University of Tokyo, Tokyo 113-8656, Japan*
[3] *Advanced Institute for Materials Research (AIMR), Tohoku University, Sendai 980-8577, Japan*
[4] *International Center for Materials Nanoarchitectonics (MANA), National Institute for Materials Science, Tsukuba 305-0047, Japan*
[5] *Research Center for Functional Materials, National Institute for Materials Science, Tsukuba 305-0047, Japan*
[6] *Research Network and Facility Services Division, National Institute for Materials Science, Tsukuba 305-0047, Japan*
[7] *Research Center for Electronic and Optical Materials, National Institute for Materials Science, Tsukuba 305-0044, Japan*
[8] *Center for Support to Research and Education Activities, Kobe University, Kobe 657-8501, Japan*
[9] *Molecular Photoscience Research Center, Kobe University, Kobe 657-8501, Japan*

**ABSTRACT**. The emergence of global phase coherence from disordered superconductors has been an issue of long-standing interest. However, disorder-induced superconductivity is observed in limited conditions and its origin is still elusive. Recently, we have reported that a highly disordered fractal MgO/MgB$_2$ nanocomposite exhibits bulk-like superconducting properties with isotropic pinning, showing an excellent phase-coherent capability irrespective of the low volume fraction (~30 vol. %) of MgB$_2$ [Uchino *et al*., Phys. Rev. B **101**, 035146 (2020); Teramachi *et al*,, Phys. Rev. B **108**, 155146 (2023)]. Hence, this nanocomposite provides a useful experimental system to investigate the relationship between the structural disorder and the establishment of the superconducting phase coherence. In this work, we show from 3D focused ion beam scanning electron microscopy (FIB-SEM) data that in the nanocomposite, a complex MgO/MgB$_2$ microstructure spreads isotropically throughout the sample with a constant fractal dimension of ~1.67. Atomic-resolution scanning transmission electron microscopy (STEM) has revealed that the MgO/MgB$_2$ interfaces are atomically clean and free from amorphous grain boundaries, even leading to atomically coherent interfaces. Detailed ac susceptibility measurements have demonstrated a smooth crossover from an intragranular to an intergranular superconducting regime, giving evidence of the establishment of the critical state due to strong intergranular coupling just below the superconducting transition temperature. Also, spatially-resolved cathodoluminescence measurements have demonstrated that oxygen vacancies in the MgO-rich phase tend to aggregate near the MgO/MgB$_2$ boundary regions, forming long channels of oxygen vacancies through the nanocomposite. These channels of oxygen vacancies will contribute to the long-range carrier transfer and the related Andreev reflection via coherent tunneling of charge carriers among the oxygen vacancy sites. Our results imply that the fractal-like MgO/MgB$_2$ microstructure with atomically clean interfaces will induce the hierarchical quantum interference of Andreev quasiparticles due to the phase coherent transport of charge carries in the MgO-rich regions, resulting in the bulk-like superconductivity in this highly disordered granular system.

---

[*] Contact author: uchino@kobe-u.ac.jp

# I. INTRODUCTION.

In the physical and materials sciences, an interface, which is defined as the boundary between two different materials or physical states, plays a vital role in controlling and understanding the properties of a material of interest. Specially designed interfaces can induce a wide range of phenomena such as mechanical strengthening, ionic conduction, magnetism, ferroelectricity and superconductivity [1–6]. Among other interfaces, a normal-superconductor (NS) interface has been a subject of considerable interest as it provides an interesting experimental and theoretical platform to study superconducting proximity effect [7–12], which is manifested as the generation of superconducting-like properties in a normal region attached to a superconductor. Microscopically, proximity effect is mediated by Andreev reflections, where incident electrons are converted into holes in the normal metal creating Cooper pairs in the superconductor [13,14]. Note also that at the interface between a normal fluid and a superfluid state of bosons, retroreflection processes analogous to Andreev reflections have been proposed to occur [15,16]. Furthermore, in the field of astronomical physics, Andreev reflections have attracted recent interest as a potential mechanism to resolve major paradoxes pertaining to the quantum description of black holes [17,18], which can behave as a superconducting surface that facilitates Andreev reflections forming Cooper pairs. Thus, the interface responsible for the Andreev reflection continues to be a topic of intense research interest in various fields of physics.

Recently, we [19,20] have shown that specially synthesized MgO/MgB$_2$ nanocomposites with ~30 vol.% of MgB$_2$ act as a fully phase-coherent bulk-like superconductor. Zero resistivity, perfect diamagnetism, isotropic pinning, and strong superfluid phase stiffness were confirmed by various experimental techniques, including electrical and magnetic measurements, transverse-field muon spin rotation (TF-$\mu$SR) spectroscopy, and magneto-optical (MO) imaging. We [19,20] have also demonstrated from electron microscopy imaging techniques that these nanocomposites are characterized not only by highly disordered microstructures but also by scale-free (or fractal) distribution of MgB$_2$ components. Since the volume fraction of superconducting MgB$_2$ components is rather low (less than ~30 vol.%), the observed bulk-like superconducting properties are difficult to understand in terms of a simple proximity-effect model. The interfacial and morphological characteristics most likely play a vital role in showing the unusual proximity effect observed in this disordered fractal system.

To get more detailed insight into the interfacial structure and the related proximity induced superconducting properties of the MgO/MgB$_2$ fractal nanocomposite, we here perform further structural analyses using focused ion beam scanning electron microscopy (FIB-SEM) and scanning transmission electron microscope (STEM) imaging. The 2D observation of FIB-SEM cross-sectional images allows us to reconstruct 3D images of the original sample, which can enable characterization of fractality of the sample over its entire volume via the resulting 3D images. High-angle annular dark field STEM (HAADF-STEM) and annular bright-field STEM (ABF-STEM) images have unambiguously demonstrated that the MgO/MgB$_2$ boundaries are characterized by atomically clean and sharp interfaces. Furthermore, we carried out a series of ac susceptibility measurements as well as detailed MO imaging analysis to investigate the flux penetration and pinning of the sample. These experimental findings reveal unique roles of interfacial and morphological characteristics responsible for the realization of the excellent phase-coherent capability of the present fractal nanocomposites. Finally, on the basis of the results of chathodeluminescence (CL) measurements, we discuss a possible role of oxygen vacancies in the MgO-rich region in terms of the charge-transport properties in the normal phase and the related Andreev reflection process.

# II. EXPERIMENTAL PROCEDURES

The MgO/MgB$_2$ fractal nanocomposites were prepared by the solid-phase reaction of Mg and B$_2$O$_3$ powders under Ar atmosphere at 700 °C, followed by a subsequent spark plasma sintering (SPS) procedure at ~1200 °C under vacuum, as reported in our previous papers [19,20]. The resulting bulk samples were cut into appropriate shapes depending on the measurement techniques, including x-ray diffraction (XRD), SEM/energy-dispersive x-ray spectroscopy (SEM-EDX), FIB-SEM, STEM, magnetoresistivity, ac and dc magnetization, MO imaging, and CL spectroscopy. Details of the sample preparation and characterization procedures are given in Supplemental Material [21].

# III. RESULTS
## A. Structural and morphological properties

Figure 1 shows a typical XRD pattern of the solid sintered sample after SPS at ~1200 °C. From the XRD pattern and the Rietveld pattern fitting, we found that the sample consists of MgO (79.2 wt %), MgB$_2$ (20.6 wt %) and a small amount of Mg (0.2 wt %), corresponding to the approximate volume fraction of MgO 73.1 %, MgB$_2$ 26.5 %, and Mg 0.4 %. The morphology and elemental distribution of the Ga-ion etched surface was investigated by SEM-EDX analysis, as shown in Fig. 2. From the SEM image given in Fig. 2(a), one sees gray and black regions, which are characterized by a complicated interwove-like structure. The SEM-EDX mapping images revealed that the oxygen (green) and boron (red) distributions almost match with the gray and black regions in the corresponding SEM image. Hence, it is reasonable to assume that the gray and black region in the SEM image correspond to the MgO- and $MgB_2$-rich regions, respectively. The box-counting fractal dimension $D$ obtained from the SEM image is 1.67 [Fig. 2(c)], confirming the fractal nature of the $MgO/MgB_2$ distribution. We then conducted the 3D reconstruction of the $MgO/MgB_2$ distribution from the sequential FIB-SEM images. The resulting 3D and cross sectional images are illustrated in Fig. 3 (see also Movie S1 in Supplemental Material [21]). We found that the disordered microstructures are isotropically distributed, yielding almost constant values of $D$ ($D$ = 1.67−1.68) irrespective of the direction of the cross section employed. This indicates that the fractal and interwoven-like $MgO/MgB_2$ structures spread isotropically throughout the sample. In this work, the bulk sintered sample was prepared by the SPS process with a uniaxial pressure of 114 MPa. From the observed isotropic morphology, we can say that the uniaxial compression during the SPS process hardly affects the resultant spatial distribution and local morphology of the MgO and $MgB_2$ grains.

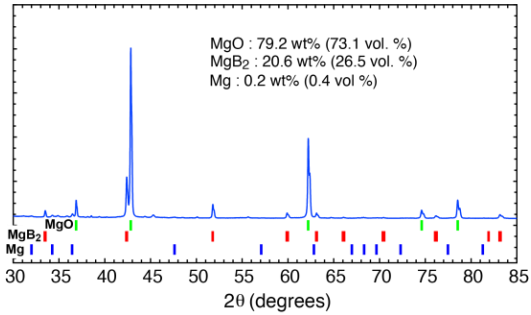

FIG.1. XRD pattern and output from quantitative Rietveld analysis of the bulk sintered sample prepared with SPS at a temperature of ~1200 ºC.

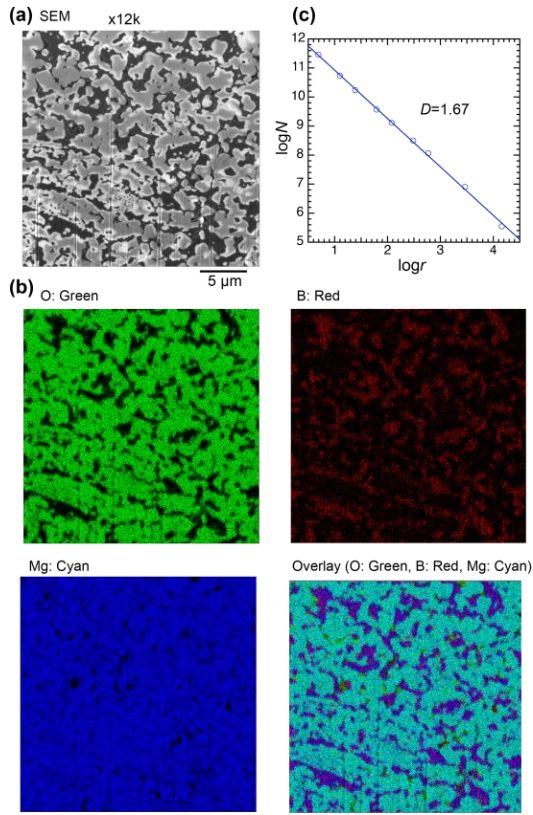

FIG. 2. (a) SEM and (b) the corresponding EDX images of the Ga-ion etched surface; Green = O, Red = B, and Blue = Mg. (c) The box-counting analysis for boron distribution in the SEM/EDX image given in (b).

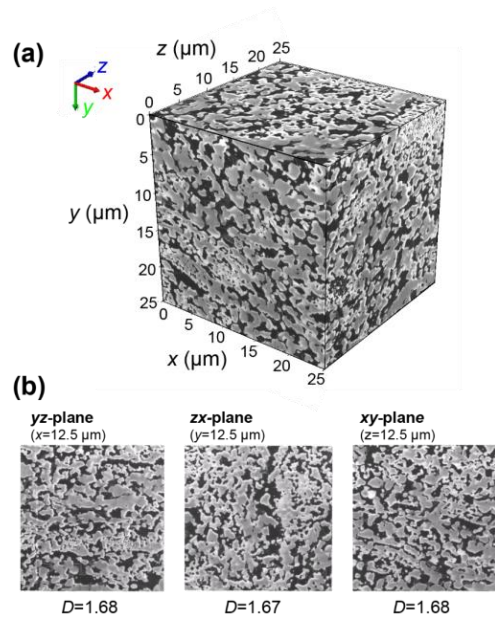

FIG. 3. (a) 3D image reconstructed from the FIB/SEM serial sectioning images. (b) Cross-sectional images sliced at $x$=12.5, y=12.5 and z=12.5 μm along the $yz$, the $zx$ and the $xy$ planes, respectively, obtained from the 3D reconstructed image. The box-counting fractal dimension $D$ is given for each image.

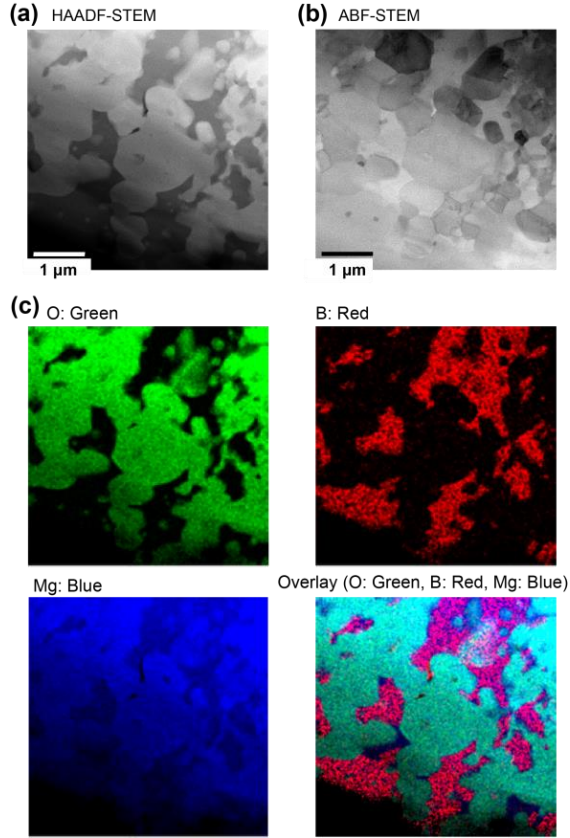

FIG. 4. (a) HAADF-STEM and (b) ABF-STEM images and (c) the corresponding STEM/EDX images; Green = O, Red = B, and Blue = Mg.

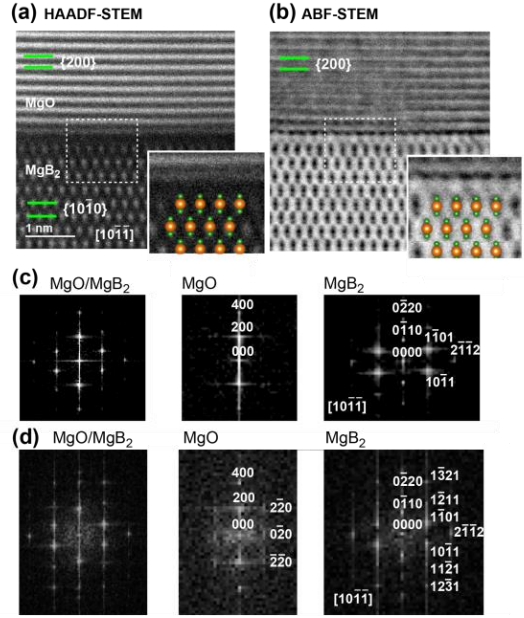

FIG. 5. (a) HAADF-STEM and (b) ABF-STEM images of a planar MgO/MgB$_2$ interface. The insets in (a) and (b) indicate the enlarged images of the white box regions, along with possible atomic arrangements of Mg (orange) and B (green) atoms. (c), (d) FFT images of (a) and (b), respectively. In (c) and (d), the left panel shows the FFT image of the whole STEM image, whereas the middle and right panels are the FFT images of the MgO- and MgB$_2$-rich regions, respectively, of the corresponding STEM images.

We next investigate the structural characteristics of the nanocomposite in the micro- and nanometer ranges using STEM. Figure 4 shows high-angle annular dark-field STEM (HAADF-STEM) and annular bright-field STEM (ABF-STEM) images along with the STEM-EDX elemental mapping. As in the case of SEM and SEM-EDX images shown in Fig. 2, the complex or fractal-like structures can be recognized in these STEM and STEM-EDX images of higher magnifications. It should also be noted that the MgB$_2$-rich regions, which are represented by red irregular shapes in the STEM-EDX image shown in Fig. 4(c), are not physically attached to each other but are separated by the surrounding MgO-rich regions. Thus, in the submicrometer range, the present nanocomposite can be regarded as 3D ensembles of randomly interconnected MgB$_2$/MgO/MgB$_2$ junctions.

Figures 5 and 6 show HAADF-STEM and ABF-STEM images of planar (straight) MgO/MgB$_2$ boundaries. In Fig. 5, the $\{10\bar{1}0\}$-type planes of MgB$_2$ and the $\{100\}$-type planes of MgO are aligned in the same direction, as also demonstrated in the corresponding fast Fourier transform (FFT) images.

The MgO/MgB$_2$ interface is atomically clean and flat without forming an interfacial amorphous phase, although the amorphous region is very often observed in the grain boundaries of polycrystalline materials [32,33]. We also found another type of clean and planar interface (Fig. 6), where the crystallographic directions of the MgO and MgB$_2$ phases appear to be different to each other. We note, however, that a close look at the FFT images reveal the alignment of the $\{110\}$-type planes of MgO and the $\{11\bar{2}2\}$-type planes of MgB$_2$ (see also the inset of Fig. 6(a)), forming a pseudo-semicoherent interface [33].

It should further be mentioned that in the nanocomposite, there exist several non-planar interfaces, as shown in Figs. 7 and 8. In the STEM images given in Fig. 7, the MgB$_2$ phase is oriented in the $[10\bar{1}\bar{1}]$ zone axis and the MgO/MgB$_2$ interface is not straight. Even in such a case, one sees that an atomically clean boundary is established by forming a terrace-and-step structure at the MgO/MgB$_2$ interface. The steps were identified to be parallel to the $\{10\bar{1}1\}$-type planes such as $(10\bar{1}1)$ and $(1\bar{1}01)$ planes of MgB$_2$ (see the insets of Fig. 7(a),(b)). Similar atomically clean MgO/MgB$_2$ interfaces with a terrace-and-step morphology can also be found in the triple junction consisting of one MgB$_2$ and two differently aligned MgO phases (Fig. 8). In this triple junction, all the constituent boundaries are free of amorphous phase as well. A terrace-and-step structure, which is along low-index planes of MgB$_2$, including ($1\bar{1}01$) and ($0\bar{1}10$) planes (see also the inset of Fig. 8(a),(b)), can be recognized at the interface between the MgB$_2$ and MgO(2) phases. The general formation of these terrace-and-step structures at the MgO/MgB$_2$ interface implies the motion and/or dynamic adjustment of grain boundaries [34–38] during the SPS process. We hence suggest that in the grain boundaries, dynamic structural transformation accommodates the SPS-induced microstructural evolution, resulting in atomically clean MgO/MgB$_2$ interfaces without creating interfacial amorphous regions.

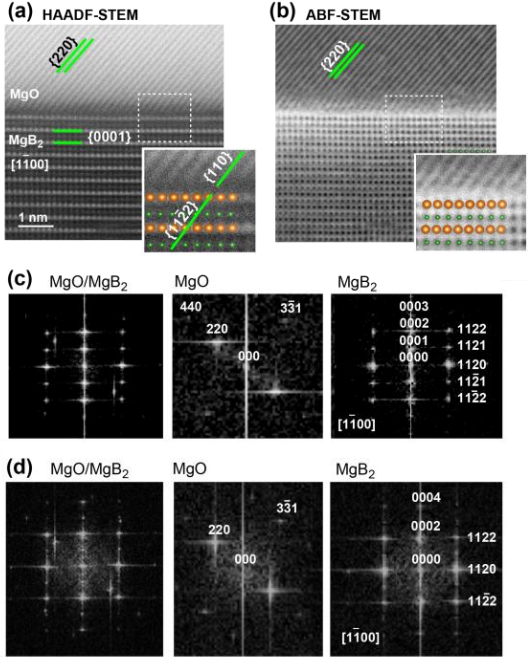
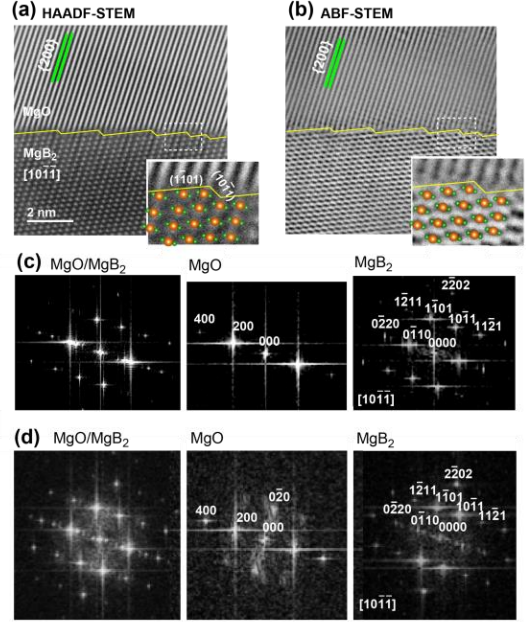

FIG. 6. (a) HAADF-STEM and (b) ABF-STEM images of another planar MgO/MgB$_2$ interface. The insets in (a) and (b) indicate the enlarged images of the white box regions, along with possible atomic arrangements of Mg (orange) and B (green) atoms. (c), (d) FFT images of (a) and (b), respectively. In (c) and (d), the left panel shows the FFT image of the whole STEM image, whereas the middle and right panels are the FFT images of the MgO- and MgB$_2$-rich regions, respectively, of the corresponding STEM images.

FIG. 7. (a) HAADF-STEM and (b) ABF-STEM images of a terraced MgO/MgB$_2$ interface. Nanometer length-scale steps are visible (marked as a solid yellow line). The insets in (a) and (b) indicate the enlarged images of the white box regions, along with possible atomic arrangements of Mg (orange) and B (green) atoms. (c), (d) FFT images of (a) and (b), respectively. In (c) and (d), the left panel shows the FFT image of the whole STEM image, whereas the middle and right panels are the FFT images of the MgO- and MgB$_2$-rich regions, respectively, of the corresponding STEM images.

The above FIB-SEM and STEM investigations on the MgO/MgB$_2$ nanocomposite not only demonstrated the isotropic fractal structure, but they also revealed the formation of atomically clean interfaces between MgO and MgB$_2$ phases. It is most likely that these unique structural and morphological characteristics are reflected in the Andreev reflection process and the related superconducting proximity effect, as has already been inferred in our previous studies [19,20]. In this work, we will further investigate their superconducting properties using the ac magnetic susceptibility technique, which provides a detailed insight about the complex electrodynamic phenomena in superconductors. Before going into details of the ac susceptibility measurements, we will briefly show the results of the resistivity $\rho$ and dc magnetization $M$ measurements, along with the distribution of magnetic flux density obtained by using the MO imaging technique.

### B. Basic superconducting properties

Figure 9(a) shows the temperature dependence of $\rho$ measured in the temperature range from 2 to 300 K. One sees that the onset of superconductivity at 38.4 K, followed by the zero-resistance state at temperatures below 36.6 K. A similar superconducting transition can also be recognized in the zero-filed-cooling (ZFC) and field-cooling (FC) dc magnetic susceptibility

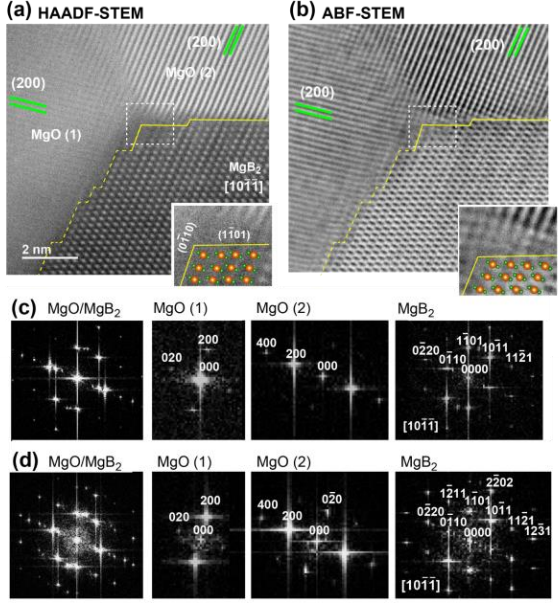

FIG. 8. (a) HAADF-STEM and (b) ABF-STEM images of a triple junction consisting of one $MgB_2$ and two differently aligned MgO phases, namely MgO(1) and MgO(2). Nanometer length-scale steps are visible especially between $MgB_2$ and MgO(2) (marked as a solid yellow line). As for the boundary between $MgB_2$ and MgO(1), the specimen is slightly tilted from the edge-on condition, yielding a somewhat blurred interface (marked as a broken yellow line). The insets in (a) and (b) indicate the enlarged images of the white box regions, along with possible atomic arrangements of Mg (orange) and B (green) atoms. (c), (d) FFT images of (a) and (b), respectively. In (c) and (d), the left panel shows the FFT image of the whole STEM image, whereas the two middle and right panels are the FFT images of the MgO(1), MgO(2) and $MgB_2$-rich regions, respectively, of the corresponding STEM images.

($4\pi\chi$) curves [Fig. 9(b)], demonstrating a superconducting onset at 38.5 K. The ZFC susceptibility curve is decreased sharply below the onset temperature and shows an almost constant value of −1.1 at temperatures below ~35 K. We should note that in this work, the dc magnetization measurements were carried out for a square cuboid-shape sample with a dimension of 1×1×5 mm$^3$ by applying a magnetic field $H$ along the long side of the sample. In this experimental set up, the effective demagnetization factor $D$ is estimated to be ~0.05 [23]. Thus, a slight discrepancy from the ideal perfect diamagnetism ($4\pi\chi=−1$) is due to a possible demagnetization effect. On the other hand, the FC curve shows a very low Meissner fraction (<~1%), illustrating the strong pinning nature of the present $MgO/MgB_2$ nanocomposite. The temperature-dependence of $\rho$ and $M$ shown above are in good agreement with those reported previously for the similarly prepared $MgO/MgB_2$ sample [20], demonstrating an excellent reproducibility of the superconducting properties of these fractal nanocomposite.

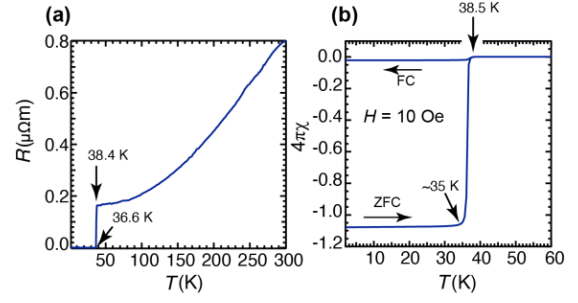

FIG. 9. (a) Temperature-dependent resistivity. (b) Zero-field-cooling (ZFC) and field-cooling (FC) magnetic susceptibility ($4\pi\chi$) curves under dc applied field of 10 Oe.

The temperature dependence of the lower critical fields $H_{c1}(T)$ was evaluated from the initial $M(H)$ ZFC curves shown in Fig. S1 in Ref. [21]. From the extrapolated value at zero temperature, $H_{c1}(0)$ is estimated to be 713 Oe (see Fig. S1(c) in Ref. [21]). The upper critical field $H_{c2}$ was obtained from the magnetoresistivity measurements (see Fig. S2 in Ref. [21]). In this work, $H_{c2}(T)$ was defined as the applied field for which the sample resistance measured at $T$ is 10 % of the normal state value. The $H_{c2}(T)$ curve was fitted with $H_{c2}(T) = H_{c2}(0)(1 - T/T_c)^{1+\alpha}$, where the parameter α > 0 describes the positive curvature of $H_{c2}(T)$ (see Fig. S2(b) in Ref. [21]). The fitted values of α and $H_{c2}(0)$ are found to be 0.23 and 97.9 kOe, respectively. From the values of $H_{c1}(0)$ = 713 Oe and $H_{c2}(0)$ = 97.9 kOe, the penetration depth λ and the coherence length ξ can be estimated to be 77.7 and 5.7 nm, respectively, on the basis of the Ginzburg-Landau (GL) theory (for the details of calculations, see Supplemental Material [21]). The obtained values of λ and ξ are also in reasonable agreement with those of our previous sample [20].

### C. MO imaging measurements

In our previous paper [20], MO imaging was used to visualize magnetic flux in the $MgO/MgB_2$ nanocomposite; however, our previous sample used for MO imaging has a physical crack created accidentally during surface polishing. In this work, we carefully prepared a crack-free plane-parallel-plate sample with a dimension of 4.1×4.8×0.5 mm$^3$ for the purpose of detailed MO analysis.

In the MO imaging, the sample was first zero-field cooled to a measurement temperature from a temperature ($T$ = ~60 K) well above the super-conducting transition temperature, and a series of MO images was acquired as the magnetic field was applied perpendicular to the sample surface up to 400 Oe (Fig. 10 (a) and the left panels in Fig. S3 [21]). One sees from Fig. 10 (a) that at a temperature of 10 K, magnetic flux hardly penetrates into the sample on applying $H$ up to 400 Oe except just along the edge, in agreement with high $H_{c1}$ values at such a low temperature range ($H_{c1}(T) > \sim 500$ Oe for $T < \sim 10$ K, see Fig. S1 [21]). As the temperature is increased, one can recognize a partial flux penetration by applying magnetic fields up to 400 Oe (Fig. 10 (a) and the left panels in Fig. S3 [21]). Note also that at temperatures up to ~33 K, the sand-pile like behavior of the perpendicular magnetic flux density $B_z$ can be seen during the ramp-up stage of $H$; i.e., the flux gradient is almost constant irrespective of $H$, in agreement with the Bean critical state model [39]. At temperatures near ~35 K, however, the profile of $B_z$ becomes flat with increasing $H$. This indicates that in these temperature and field regions, the magnetic flux gradient along the in-plane ($x$-axis) direction should decrease with $H$ and that the corresponding current density $J_c$ should diminish, as predicted by a modified critical state model assuming that the critical current density is not constant, but is determined by the pinning force [40,41].

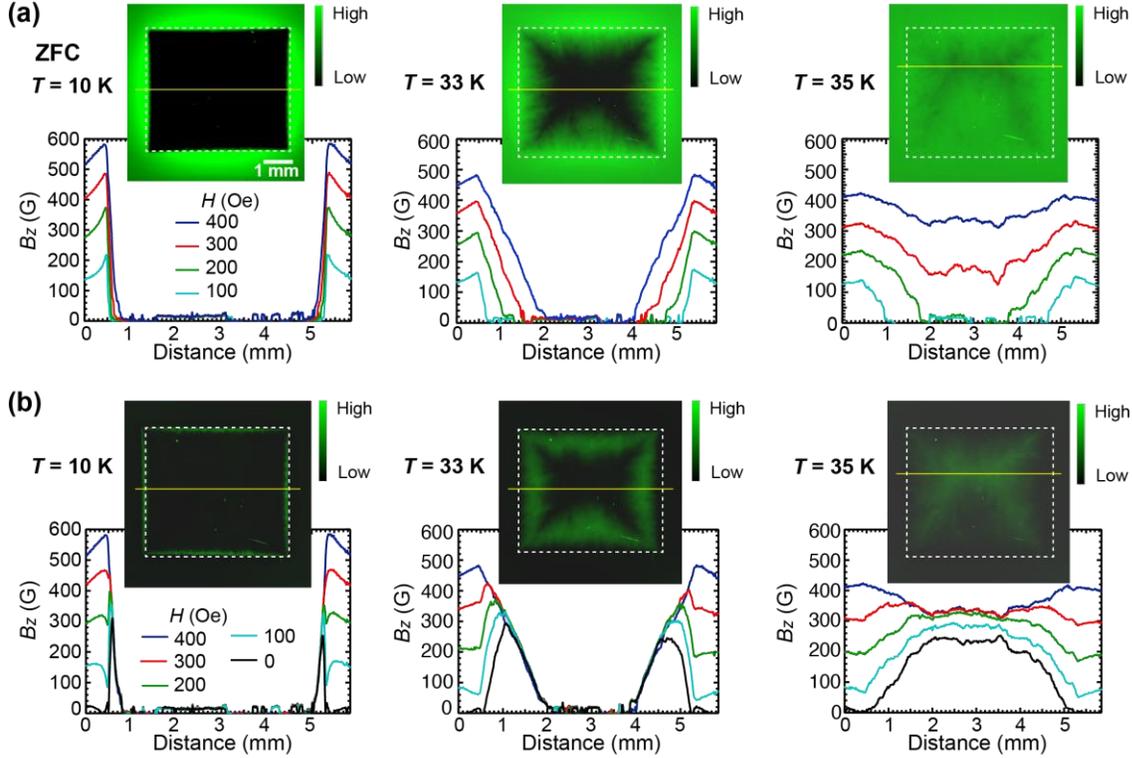

FIG. 10. MO observations after ZFC to different temperatures. (a) Profiles of perpendicular magnetic flux density $B_z$ along the yellow line of the inset during the ramp-up stage of $H$ from 100 to 400 Oe. The respective insets show the MO images obtained in an applied field of 400 Oe. (b) Profiles of $B_z$ along the yellow line of the inset during the ramp-down stage of $H$ from 400 to 0 Oe. The respective insets show the MO images in the remanent ($H = 0$ Oe) state showing trapped magnetic flux.

We next investigate the flux density profile during the ramp-down stage of $H$. As shown in Fig. 10 (b) and the right panels Figs. S3 [21], the observed $B_z$ profiles obtained at temperatures up to ~33 K follow the critical state model, in agreement with the results obtained in the ramp-up stage of $H$. Note also that at temperatures around ~35 K, where the vortices occupied the whole crystal for $H \geq \sim 300$ Oe, the vortices around the sample edges leave the sample with decreasing $H$, and accordingly, in the remanent state, the penetrated vortices in the near central region of the sample are strongly pinned to show a "rooftop" pattern [see the right panel in Fig. 10(b)]. In the remanent state at a temperature of 35 K, however, the ridge of the rooftop is not located in the exact middle of the sample, implying that the pinning force density is not constant over the entire volume of the sample at such a temperature close to $T_c$.

We should note that $B_z$ obtained from the MO observations of a thick-plate sample is generated mainly by the sum of a two-dimensional vortex current flow in the near-surface region [30]. From this follows that a numerical inversion of Biot-Savart law can yield information on the semiquantitative current distribution in the sample surface [29] (for details, see Supplemental Material [21]). Figure 11 shows mapping of the current density and the current vectors calculated by using the corresponding MO images shown in Fig. 10. When a magnetic field is applied after ZFC, the current flows in a clockwise direction throughout the sample, implying the presence of uniform bulk current in the corresponding critical state. As for the remanent states, one sees complex current distributions flowing in a clockwise and/or an anti-clockwise direction depending on the measurement temperature. The resulting complex current flow in the remanent state is due to the redistribution of the pinned vortices during the field ramp down process and the subsequent local change in the sign of the gradient $\frac{\partial B_z}{\partial x}$, as shown in Fig. 10 (b). Thus, the current distributions provide complementary and useful information on the temperature- and field-dependent flux pinning of the present sample.

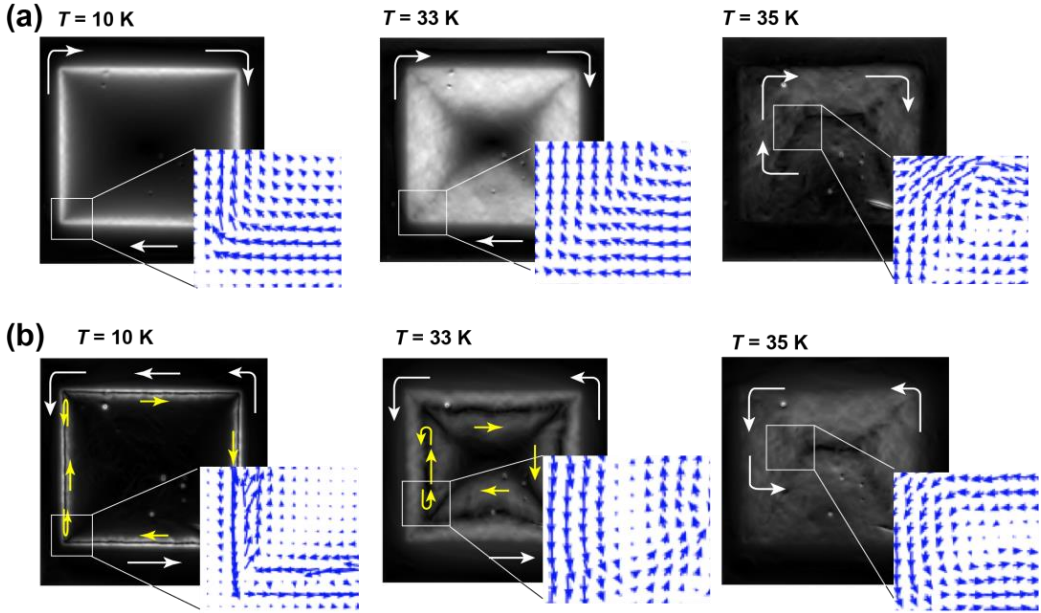

FIG. 11. (a), (b) Supercurrent density calculated from the MO images shown in the insets of Figs. 10(a), and 10(b), respectively. The white and yellow arrows qualitatively indicate the direction of the local current. The respective insets represent the calculated current of the white box region in the form of current vectors.

Furthermore, we measured the MO image of trapped magnetic flux in the field-cooled (FC) remanent state, which is produced by cooling the sample from ~60 to 5 K under an external magnetic field of 200 Oe and then reducing the applied field to zero. Figures 12(a) shows the MO image for the remanent state taken at 5 K, and its qualitative 3D representation is given in Fig. 12(c). We see that the magnetic flux is distributed uniformly throughout the sample, as evidenced by an almost flat profile of $B_z$~200 G (see the inset of Fig. 12(a)). This indicates that the vortices are strongly and uniformly pinned over the whole area of the sample at such a low temperature. Figures 12 (a) and 12(c) also revealed that the flux distribution near the edge is steeply reduced, showing a flux annihilation zone just outside the edge. Figure 12(b) gives the corresponding current distribution. One sees that a large supercurrent circulates especially around the sample edge in an anti-clockwise direction. These edge features most likely results from the uniform return field of the trapped flux [42,43]. The flux of the opposite sign penetrates an outer rim of the sample, leading to a steep drop in $B_z$ and the corresponding edge supercurrent, together with the flux annihilation zone outside the edge.

### C. AC susceptibility measurements

From a series of MO images demonstrated in the above subsection, one can notice that the present MgO/MgB$_2$ nanocomposite does not show any granular behaviors in terms of flux entry, exit and pinning, but it exhibits good electromagnetic homogeneity at least at temperatures below ~36 K. To get an additional insight into the superconducting properties in the temperature region above ~36 K, we employed the ac magnetic susceptibility method, which is a useful technique for the analysis of the electromagnetic properties of various forms of superconductors (single crystals, polycrystals, films, etc), especially under weak magnetic fields just below $T_c$.

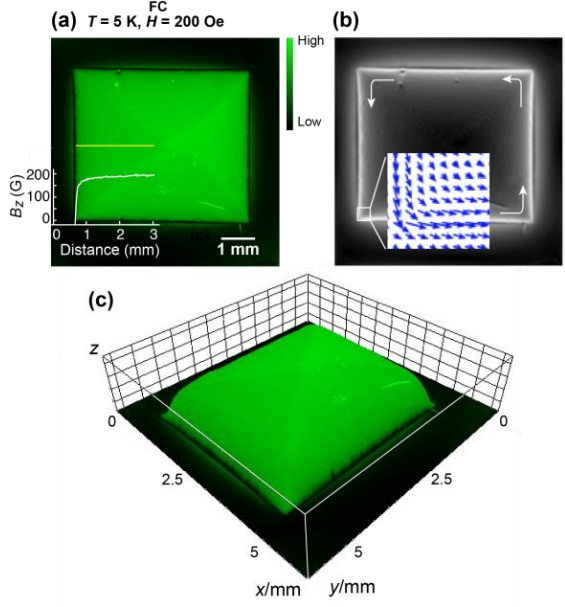

FIG. 12. (a) MO image after FC to 5 K in an applied field of 200 Oe, and (b) the corresponding calculated current density. The inset in (a) represents the profile of perpendicular magnetic flux density $B_z$ along the yellow line. The inset in (b) shows the calculated current of the white box region in the form of current vectors. The white arrows in (b) qualitatively indicate the direction of the local current. (c) Qualitative 3D representation of the MO image shown in (a).

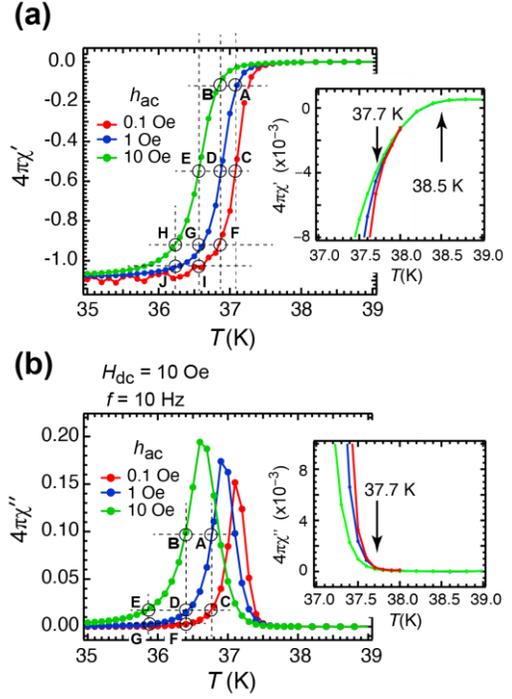

FIG. 13. Temperature dependence of the magnetic ac susceptibility [(a) $4\pi\chi'$ and (b) $4\pi\chi''$] observed in a dc magnetic field of 10 Oe using an ac frequency $f$ of 10 Hz at three different ac amplitudes $h_{ac}$. The respective insets show the expanded plots near the superconducting onset temperature ($T$ = 38.5 K).

Figure 13 shows the temperature-dependent ac susceptibility of the MgO/MgB$_2$ nanocomposite with a dimension of 1×1×5 mm$^3$ measured in a static (dc) magnetic field of $H$ = 10 Oe with different ac excitation amplitudes $h_{ac}$ and a constant frequency $f$ of 10 Hz. The magnetic field is applied along the long side of the sample, as in the case of the dc susceptibility measurements. One sees from Fig. 13(a) that the real ($4\pi\chi'$) susceptibility shows an onset of superconductivity at 38.5 K and reaches ∼−1 at $T$∼35 K, in agreement with those obtained from the dc magnetization measurements shown in Fig. 9(b). In the temperature range from 37.7 to 38.5 K, the $4\pi\chi'$ curves are almost independent of $h_{AC}$, and the imaginary ($4\pi\chi''$) susceptibility remains to be zero. This $h_{AC}$-independent feature of the ac susceptibility can be interpreted in terms of the intragranular superconductivity [44,45]. At temperatures below 37.7 K, both the real and imaginary parts of the ac susceptibility show a strong $h_{AC}$ dependence. The resulting large decrease of $4\pi\chi'$ as well as the dissipative peak in $4\pi\chi''$ results from the intergranular superconducting properties [44,45]. The observed strong $h_{AC}$-dependence of the ac susceptibility in the regime of intergranular superconductivity is a consequence of the nonlinearity in the electromagnetic response and can be described within the framework of the critical state model [39].

To further prove that the critical state model is applicable to the present ac susceptibility data, we employed a geometrical test proposed by Civale *et al.* [46]. In the critical state for a given sample, $4\pi\chi'$ and $4\pi\chi''$ is a only a function of ac penetration length $D_c = \frac{c}{4\pi}\left(\frac{h_{ac}}{J_c}\right)$, where $c$ is the velocity of light, and beyond $D_c$ the ac field is totally screened. Consequently, every horizontal line for the $4\pi\chi'$ and $4\pi\chi''$ curves shown in Fig. 13 corresponds to a constant value of $D_c$. On the other hand, every vertical line corresponds to a constant value of $J_c$ as $J_c$ is only a function of temperature. This implies that $J_c(A)=J_c(C)$, where A ($h_{ac}$= 1 Oe) and C ($h_{ac}$= 0.1 Oe) are points located in the same vertical line in Figs. 13(a) and 13(b). The point B, with the same $D_c$ as point A, belongs to the curve $h_{ac}$ = 10 Oe. Since the critical state model establishes that $J_c \propto h_{ac}/D_c$, we get $J_c(B)=10J_c(A)$. Similarly, the point D of the curve of $h_{ac}$= 1 Oe has the same $D_c$ as the point C, which is the curve of $h_{ac}$= 0.1 Oe, and hence $J_c(D)=10J_c(C)$. The consequence is that $J_c(B)=J_c(D)$, implying that the points B and D should be located in the same vertical line (or the same $T$), as indeed graphically confirmed in both the $4\pi\chi'$ and $4\pi\chi''$ curves shown in Figs. 13(a) and 13(b). The occurrence of other rectangles, such as DEFG and/or GHIJ in the $4\pi\chi''$ and/or $4\pi\chi'$ curves, further confirms the validity of the critical state model.

The above geometrical construction allows us to determine the temperature dependence of $J_c$ in the temperature region just below $T_c$ [46]. If we define that $J_c(D)=1$ (in arbitrary units) in the $4\pi\chi'$ curve of $h_{AC}$= 1 Oe, we find $J_c(E) =10J_c(D)$ and $J_c(C) =0.1J_c(D)$ according to the above argument. One also notices from the $4\pi\chi'$ curves shown in Fig. 13(a) that the points C, D, and E roughly represent the temperature of half screening $T_{hs}$ under the respective ac excitation field amplitudes, i.e., $T_{hs}$ = 37.1 K (C), 36.9 K (D) and 36.6 K(E) for $h_{AC}$= 0.1, 1, 10 Oe, respectively. Then the absolute values of $J_c$(C). $J_c$(D) and $J_c$(E) can be estimated to be 1.6, 16 and 160 A/cm$^2$ by assuming that for our geometry, $D_c$=0.5 mm, i.e, a half thickness of the sample. As shown in Fig. 14, the thus obtained $J_c$ values show a highly concave temperature dependence, which can be tentatively fitted with $J_c \propto (1-T/T_c)^{3/2}$ in the temperature range observed here. According to Clem et al. [47], $J_c$ obeys the concave Ginzburg–Landau (GL) $(1-T/T_c)^{3/2}$ temperature dependence in a granular superconductor in which the ratio of the Josephson-coupling energy to the superconducting condensation energy is of unity or larger. It can hence be expected that the present sample behaves as a strongly coupled homogeneous (or spatially isotropic) granular superconductor when the system is cooled below ~37 K. The inset of Fig. 14 also includes the $J_c$ values at lower-temperature ($T \leq$ 30 K) derived from the height of the magnetization loop at zero magnetic field (see Fig. S4 [21]) on the basis of the Bean model [39] (for details see Supplemental Material [21]). The measured $J_c$ data in the full (2–~37 K) temperature range reasonably follows the mean-field behavior predicted by the GL theory [48]:

$$J_c(T) = J_c(0)(1-(T/T_c)^2)^{3/2}((1+(T/T_c)^2)^{1/2}, \quad (1)$$

where $J_c(0)$ is the density of the depairing critical current at 0 K. Note that near $T_c$ Eq. (1) reduces to the familiar GL $(1-T/T_c)^{3/2}$ temperature dependence [48]. The applicability of the GL theory confirms the establishment of the strongly-coupled super-conductivity at temperatures below $T$~37 K.

In the ac susceptibility measurements, the effect of the sweep frequency $f$ on the $4\pi\chi'$ and $4\pi\chi''$ curves is also worth investigating [49]. In general, the weak $f$ dependence of the ac susceptibility is related to flux creep of vortices, being a slow process, contrary to the strong $f$ dependence predicted for a linear conductor [50]. Figure 15 shows the $4\pi\chi'$ and $4\pi\chi''$ curves obtained at $h_{ac}$= 1 Oe under different sweep frequencies in zero static magnetic field. One sees that at temperatures above ~37 K, the $4\pi\chi'$ and $4\pi\chi''$ curves shift to higher temperatures and the transition width broadens with increasing $f$, also accompanied by an intensity increase in the ac loss peak. These frequency dependent features are caused by the vortex dynamic phenomena and the related dissipation processes, including conventional eddy currents, as often observed in type-II high-$T_c$ superconductor [51,52] and a pure MgB$_2$ bulk sample [53]. Note, however, that when the temperature of the system goes below ~36.7 K, such a frequency dependence is not observed both in the $4\pi\chi'$ and $4\pi\chi''$ curves. This indicates that at a temperature of ~36.7 K, the system is transformed from a linear conductor into an ideal "Bean" superconductor, in which the ac susceptibility is nonlinear and frequency independent [50]. This result is in consistent with that obtained from the $h_{ac}$ dependence given in Fig. 13. Thus we can conclude that the critical state model is applicable to the present sample at temperatures below ~37 K, which is lower than the intragranular superconducting transition temperature ($T_c$=38.5 K) only by around 1.5 K.

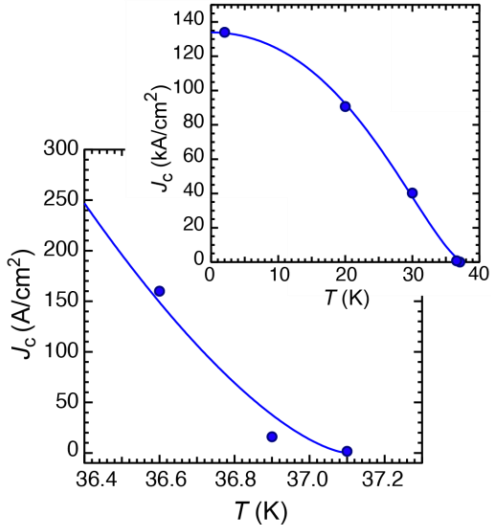

FIG. 14. Critical current density $J_c$ as a function of temperature $T$ derived from the ac susceptibility data of Fig. 13(a). Solid line represents fit of function $J_c \propto (1-T/T_c)^{3/2}$. The inset includes the lower-temperature ($T \leq 30$K) $J_c$ values derived from the height of the magnetization loop at zero magnetic field on the basis of the Bean model [21]. The solid line in the inset represents the temperature dependence of the depairing critical current density according to Eq. (1).

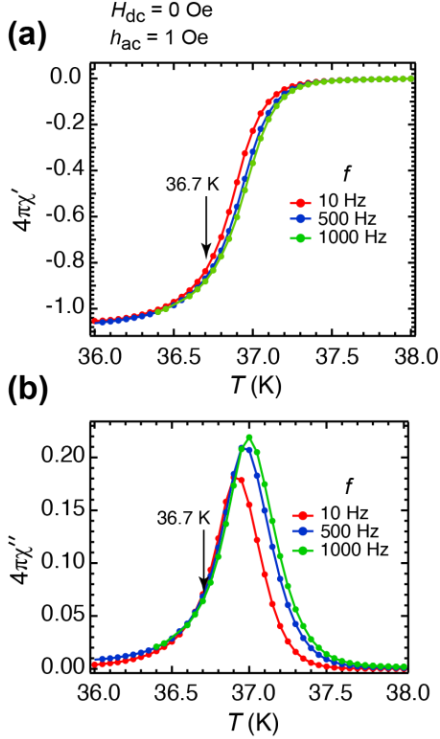

FIG. 15. Temperature dependence of the magnetic ac susceptibility [(a) $4\pi\chi'$ and (b) $4\pi\chi''$] observed in a zero dc magnetic field using an ac amplitudes $h_{ac}$ of 1 Oe and three different ac frequencies.

## IV. DISCUSSION

From the above observations, we have shown that the present MgO/MgB$_2$ nanocomposite behaves as a homogeneous bulk-like superconductor in a critical state at temperatures below ~37 K. The transition from the intra- to inter-granular regime is very smooth and is not apparently noticeable in the resistivity and susceptibility curves. These features are quite different from those of conventional granular superconductors, which generally show two-step behaviors in the temperature-dependence of resistivity and susceptibility due to the higher-temperature intragranular transition and the subsequent lower-temperature intergranular transition [54]. This two step feature is understood in terms of a weak link model [49,55,56] or a two-level critical state model [57] assuming two distinct critical densities, i.e., a critical current density inside the grains, and another one reflecting the intergranular coupling. In general, the intergranular critical current density is much smaller than the intragranular one, and hence that magnetic flux easily penetrates along the grain boundaries or weak links [58,59], showing two-step and the related inhomogeneous superconducting properties in granular superconductors. In Ref. [20], we have also confirmed from the Hall measurements that in the normal state, the net hole density in this nanocomposite is an order smaller than that of pure MgB$_2$, as expected from the low volume fraction (~30%) of MgB$_2$. Such a low carrier density would make the sample too resistive to sustain superconductivity, as indeed observed in a MgO/MgB$_2$ system with ~40 wt. % of MgO [60]. Hence, the occurrence of the bulk-like superconductivity in the present MgO/MgB$_2$ nanocomposite implies that exceptionally strong and long-range intergranular phase locking states are established in this system. Considering that a fractal topology can provide an underlying basis for the establishment of long-range correlation and cooperation behaviors in complex systems [61−63], we suggest that the fractal morphologies along with the atomically clean interfaces allow robust and hierarchical Andreev interference effects responsible for the strong and long-range phase coherence throughout the sample. The above considerations provide a possible explanation for the establishment of global phase coherence in highly disordered "granular" superconductors. This knowledge may further shed light on the emergence of global phase coherence from local pairing in high-$T_c$ cuprates [64,65] and arrays of superconducting grains [66,67]. This is because these systems are characterized by low superfluid density, meaning that their superconducting properties are governed by phase fluctuations and are very sensitive to the structural and building characteristics. The situation is very similar to the case of the present sample with a low carrier density.

Although it is certain that the MgO/MgB$_2$ fractal nanocomposite has an excellent phase-coherent capability, we still have one unanswered question. Where do the carriers in the normal phase or MgO come from? MgO is, in principle, an insulator with a large bandgap of 7.8 eV, and no carriers are present in structurally perfect MgO. It should be noted, however, that we prepared these nanocomposites in highly Mg rich conditions using the solid-phase reaction between Mg and B$_2$O$_3$ (for details, see the sample preparation section in Supplemental Material [21]). It is hence probable that large amounts of intrinsic defects such as O vacancies and Mg interstitials are present in our sample. Such intrinsic defects in MgO can participate in both the electron and hole conductivity, causing bipolar charge transport [68]. Among other intrinsic defects, single oxygen vacancies, called F centers, and/or double oxygen vacancies, called $F_2$ centers, are the most likely candidates for the intrinsic defects in MgO [69–74]. Note also that McKenna and Blumberger [75] have shown from first-principles modeling that in defective MgO with a high concentration of oxygen vacancies, coherent electron tunneling is possible to occur when the separations between oxygen vacancy defects become less than 0.6 nm, accounting for the origin of long-range carrier transfer in metal oxide materials. From these considerations, we can assume that in the MgO/$MgB_2$ nanocomposite, there exists a large number of intrinsic defects possibly in the form of F and $F_2$ centers, which will contribute to the bipolar charge transport and hence to the expected Andreev reflection.

To confirm the assumption, we performed CL measurements on the polished surface of the MgO/$MgB_2$ nanocomposite at room temperature. It has been well documented that oxygen vacancies in MgO show a variety of luminescence properties in the ultraviolet/visible spectral regions depending on their charge states and atomic configurations [70–72, 76–78]. Hence, CL spectroscopy is a powerful technique to identify the quality and quantity of possible oxygen vacancies present in our sample.

Figures 16 and 17 show SEM images of low (3000×) and high (8000×) magnifications, respectively, along with the corresponding CL spectra. As demonstrated previously, the gray region in the SEM images represent the MgO-regions, whereas the black regions correspond to the $MgB_2$-rich regions. From the position dependent CL spectra, we see that the CL emission properties vary from site to site. In the black (or $MgB_2$-rich) regions, we hardly observed any CL emissions [see the CL spectra of position A in Figs. 16(b) and 17(b)]. This is reasonable because $MgB_2$ is a metallic compound at room temperature, showing no band-gap and/or defect-related luminescence. As for the gray (or MgO-rich) regions, we typically observed two types of CL spectra. One is characterized by the CL band peaking at ~350 nm with a longer-wavelength shoulder extending to ~800 nm [see the CL spectra of position B in Figs. 16(b) and 17(b)]. The prominent CL peak at ~350 nm is assigned to the emissions from $F_2$ (375 nm) and/or $F^+$ (380 nm) centers [70–72,76], and the longer-wavelength tail to those from $F_2^{2+}$ (441 nm), F (490 nm) and $F_2^+$ (475 nm) centers [77,78]. The other type of CL spectra shows two broad CL bands at ~500 nm and ~700 nm [see the CL spectra of position C in Figs. 16(b) and 17(b)]. The CL band at ~500 nm results from the F center, with minor contributions from $F_2^{2+}$, and $F_2^+$ centers, while the one at ~700 nm is probably due to the thermal detachment of holes from the Mg vacancy [79]. To elucidate the spatial distributions of these two types of CL emissions, CL mapping images were taken at wavelengths of 350 and 520 nm, as shown in Figs. 16(c) and 17(c). Although the CL mapping images at 350 and 520 nm are not identical to each other, they are mostly overlapped, implying that F- and $F_2$-type centers spatially coexist. It should also be noted that these F- and $F_2$-type centers are not distributed homogeneously over the entire MgO-rich regions, but rather they are preferentially located around the $MgB_2$-rich regions, forming long and winding channels consisting of oxygen vacancies. These observations strongly support our assumption that the Andreev reflection at the $MgB_2$/MgO interface is assisted by bipolar (electron and hole) charge transport among the F- and $F_2$-type centers, which are located in close vicinity to the $MgB_2$ phase. The coherent carrier transport along the channel of oxygen vacancies will promote the long-range proximity effect via successive Andreev reflection processes and will eventually lead to the establishment of strong macroscopic phase coherence throughout the sample.

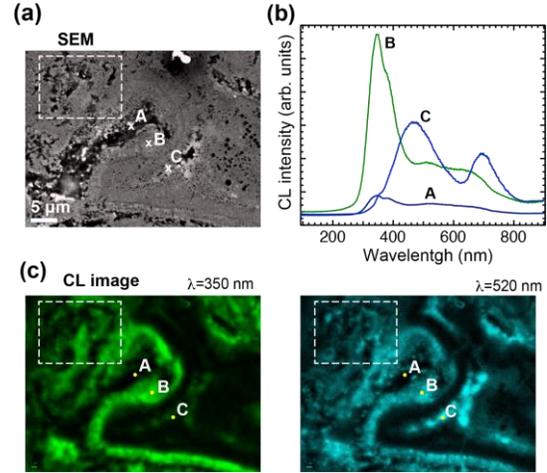

FIG. 16. (a) SEM image (3000×) and (b) the corresponding CL spectra acquired at points A, B, and C in (a). (c) CL mapping images obtained at wavelengths of 350 (left panel) and 520 (right panel) nm. White boxes in (a) and (c) indicate appropriate location used for the subsequent CL observations give in Fig. 17.

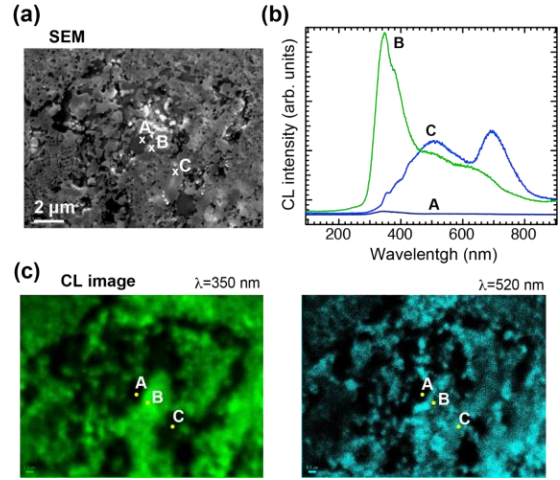

FIG. 17. (a) SEM image (8000×) and (b) the corresponding CL spectra acquired at points A, B, and C in (a). (c) CL mapping images obtained at wavelengths of 350 (left panel) and 520 (right panel) nm. These SEM and CL images are taken for the white box region given in Fig. 16.

## V. CONCLUSIONS

We have performed detailed structural and morphological characterization on the MgO/$MgB_2$ fractal nanocomposite. The FIB-SEM and the resulting 3D reconstructed images have elucidated that the randomly interwoven-like MgO/$MgB_2$ structures spread isotropically throughout the sample, showing an almost constant value of fractal dimension ($D$ = 1.67 – 1.68) irrespective of the direction of the cross section inspected. Atomic-scale STEM analysis has revealed that the atomically clean MgO/$MgB_2$ interfaces are created in the MgO/$MgB_2$ nanocomposite. The interface boundaries are characterized by planar boundaries and/or terrace-and-step structures without forming interfacial amorphous regions. These clean interfaces result presumably from the motion and atomic arrangements of grain boundaries during the sintering process. We have shown from resistivity, dc and ac susceptibility, and MO measurements that at temperatures below ~37 K, the present MgO/$MgB_2$ nanocomposite acts as a homogeneous bulk-like superconductor in a critical state, which is preceded by an intragranular superconducting transition at 38.5 K. The strong intergrain coupling is expected to be emerged over the entire volume of the system although the sample consists of MgO and $MgB_2$ nanograins with complex morphologies. Also, the SEM-CL measurements have revealed the presence of large amounts of oxygen vacancies in the MgO-rich phase, forming complex long and winding channels consisting of oxygen vacancies around the $MgB_2$-rich phase. These channels of oxygen vacancies will allow the long-range carrier transfer via coherent tunneling and hence the long-range proximity effect due to hierarchical quantum interference of Andreev quasiparticles. The strong macroscopic phase coherence in the highly disordered MgO/$MgB_2$ nanocomposite could be realized due to the fortuitous combination of atomically clean interface, fractal morphology, and long channels of coherent charge transport. These structural features presumably induce long-range proximity-induced superconducting correlations, accounting for the excellent phase-coherent capability of this proximity-coupled fractal system.

## ACKNOWLEDGMENTS

A part of this work was conducted in Institute for Molecular Science, supported by "Advanced Research Infrastructure for Materials and Nanotechnology in Japan (ARIM)" of the Ministry of Education, Culture, Sports, Science and Technology (Proposal Numbers. JPMXP1223NM0163, JPMXP1223MS1024 and JPMXP1224MS1017). This work was also supported by NIMS Joint Research Hub Program. We also acknowledge Research Facility Center for Science and Technology, Kobe University, for providing access to the MPMS facility.

## AUTHOR CONTRIBUTIONS

T. U. and I. N conceived the project. I. N. prepared the samples. S. K. and Y. I. performed TEM measurements. S. O., M. T. and S. A. conducted MO measurements. I. N., A, H. and T. U. analyzed the MO data. A. N. and T. M performed FIB-SEM measurements and 3D image reconstruction. J. C. and H. S. contributed to CL experiments. T. U., I. N., T. S. and H. O. performed dc and/or ac susceptibility measurements. The manuscript was written by T. U. with comments and input from all authors.

## DATA AVAILABILITY

The data that support the findings of this article are available from the authors upon reasonable request.

**Supplemental Material**
**Establishment of global phase coherence in a highly disordered fractal MgO/MgB$_2$ nanocomposite: Roles of interface, morphology and defect**

Iku Nakaaki,[1] Aoi Hashimoto,[1] Shun Kondo,[2] Yuichi Ikuhara,[2,3] Shuuichi Ooi,[4] Minoru Tachiki,[4] Shunichi Arisawa,[5] Akiko Nakamura,[6] Taku Moronaga,[6] Jun Chen,[7] Hiroyo Segawa,[7] Takahiro Sakurai,[8] Hitoshi Ohta,[9] and Takashi Uchino[1]

[1] *Department of Chemistry, Graduate School of Science, Kobe University, Kobe 657-8501, Japan*
[2] *Institute of Engineering Innovation, School of Engineering, The University of Tokyo, Tokyo 113-8656, Japan*
[3] *Advanced Institute for Materials Research (AIMR), Tohoku University, Sendai 980-8577, Japan*
[4] *International Center for Materials Nanoarchitectonics (MANA), National Institute for Materials Science, Tsukuba 305-0047, Japan*
[5] *Research Center for Functional Materials, National Institute for Materials Science, Tsukuba 305-0047, Japan*
[6] *Research Network and Facility Services Division, National Institute for Materials Science, Tsukuba 305-0047, Japan*
[7] *Research Center for Electronic and Optical Materials, National Institute for Materials Science, Tsukuba 305-0044, Japan*
[8] *Center for Support to Research and Education Activities, Kobe University, Kobe 657-8501, Japan*
[9] *Molecular Photoscience Research Center, Kobe University, Kobe 657-8501, Japan*

## I. METHODS
### A. Sample preparation.

The bulk Mg/MgO/MgB$_2$ nanocomposites were prepared by the solid phase reaction of Mg and B$_2$O$_3$ powders under Ar atmosphere at 700 ˚C followed by a subsequent spark plasma sintering (SPS) procedure according to the procedure reported in our previous papers [19,20]. Pure Mg (99.9%) and B$_2$O$_3$ (99.9%) powders with a molar ratio of Mg/B$_2$O$_3$=5:1 were used as starting material. The Mg/B$_2$O$_3$ mixture with a total weight of 2 g was put in a cylindrical alumina crucible. This crucible was located inside a larger rectangular alumina crucible, which was closed with a thick aluminum lid. This set of crucibles was placed in an electric furnace, which was evacuated down to ~30 Pa and was subsequently purged with argon. The temperature of the furnace was raised to 700 ˚C at a rate of ~10 ˚C/min and kept constant at 700 ˚C for 6 h under flowing Ar environment. After the heating process, the furnace was naturally cooled to room temperature, yielding black powders in the inner crucible. We confirmed from powder X-ray diffraction (XRD) measurements that the resulting powder consists of MgO, Mg, and MgB$_2$. The collected black powders weighing about 2.0 g were loaded into a 15 mm diameter graphite die and were processed using a SPS system (SPS-322HG, Fuji Electronic Ind. Co., LTD.). In SPS, sintering is realized by subjecting the green compact to arc discharge generated by a pulsed electric current. An electric discharge process takes place on a microscopic level and accelerates the sintering processes accompanied by material diffusion. The pulsed electric current (2000 A) was passed through the sample under dynamic vacuum (~50 Pa) while a 114 MPa uniaxial pressure was applied. The heating rate was ~450 ˚C/min up to 1200 ˚C. During the SPS procedure, the temperature, applied pressure, displacement (shrinkage), and environmental pressure were recorded. After the SPS treatment, a cylindrical bulk sample with a diameter of 15 mm and a length of ~4 mm was obtained. The bulk density of 3.15 g/cm$^3$, which corresponds to 97 % densification, was obtained by dividing the weight in air by the geometric volume of the specimen. The thus obtained SPS-treated samples were cut into suitable shapes for later characterization.

**B. X-ray diffraction measurements.**
X-ray diffraction (XRD) patterns of the SPS-treated sample were obtained with a diffractometer (SmartLab, Rigaku) equipped with a sealed tube X-ray generator (a copper target; operated at 40 kV and 30 mA). Rietveld refinement of XRD patterns was performed to accurately determine the phases and their quantitative compositions using Rigaku's PDXL software with the Whole Pattern Powder Fitting (WPPF) method, connected to the PDF2 database of International Centre for Diffraction Data (ICDD). The refinements were carried out by checking the fitting quality of the powder patterns by means of two parameters, i.e., the goodness-of-fit indicator GoF and the weighted R-factor of Rietveld refinement $R_{wp}$.

**C. Focused ion beam scanning electron microscopy (FIB-SEM ) observation.**
FIB and energy dispersive X-ray (EDX) spectroscopy were conducted on the sample cut from the SPS-treated samples using an FIB-SEM system (SMF-1000, HITACHI High-Tech) with an EDX spectrometer. The specimen was milled using a Ga ion beam at a 30 kV acceleration voltage and a 3 nA beam current with a milling interval of 25.0 nm. The milled surfaces of a $25 \times 25$ μm$^2$ region with a pixel size of $12.5 \times 12.5$ nm$^2$ were imaged by SEM at an acceleration voltage of 5 kV using in-lense SE detector. From this procedure, we obtained a set of 1000 FIB-SEM serial sectioning images. After aligning the data set, image stacks were segmented to reconstruct a 3D image with a voxel size of $12.5 \times 12.5 \times 25.0$ nm$^3$.

**D. Scanning transmission electron microscopy (STEM) observation.**
High-angle annular dark-field (HAADF) and annular bright-field (ABF) STEM observations and STEM-EDX mappings were performed with a scanning transmission electron microscope (JEM-ARM200F, JEOL Ltd.) equipped with an aberration corrector (ASCOR, CEOS GmbH) operated at 200 kV. For the STEM measurements, the sample was thinned using argon ion milling technique with an ion slicer (EM-09100IS, JEOL Ltd.) until perforation occurred in the middle of the sample. The convergence semi-angle angle was set to 24 mrad, and the collection angle for HAADF- and ABF-STEM observations were > 54 mrad and 12-24 mrad, respectively..

**E. Fractal analysis.**
To evaluate the fractal dimension for the distribution of MgB$_2$ nanograins in the composite, box counting was applied on SEM images and FESEM-EDX mapping images of boron Kα. Firstly the original images were converted to an 8-bit format and then finally transformed to a binary format. Image processing and fractal analysis were carried out in the ImageJ 1.52a software (NIH). Box counting technique is consisted of covering the slice image with voxels (boxes in 2D). The fractal dimension or box-counting dimension $D$ is obtained from the following relationship [22]

$$D = \frac{\log(N)}{-\log(r)}, \quad (S1)$$

where $r$ is variable box dimension in pixels and $N$ is the minimum number of boxes needed to encompass the whole object containing the boron Kα signals. If any linear region is observed in the curve log($N$) versus log($r$), the fractal dimension is equal to the slope with a negative sign of the curve.

**F. Electrical resistivity and magnetoresistivity measurements.**

The electrical resistivity under magnetic field up to 70 kOe was measured out using the standard dc four-probe technique in the 2–300 K range using the square cuboid sample with a size of 2×2×10 mm$^3$. The magnetic field was provided by superconducting magnets in a commercial superconducting quantum interference device (SQUID) magnetometer (MPMS-XL, Quantum Design, San Diego, USA). In this work, the longitudinal magnetoresistivity was measured; that is, the magnetic field was aligned along the direction of current flow..

**G. DC magnetization measurements.**

The dc magnetization measurements in external fields $H$ were carried out for a square cuboid shape sample with dimensions of 1×1×5 mm$^3$ using a commercial SQUID magnetometer (MPMS-XL, Quantum Design). The field was applied along the long side of the cuboid sample. In this configuration, the resulting effective demagnetization factor is estimated to be less than 0.05 [23], and the magnetizations and susceptibilities were not corrected for this small demagnetization factor. To estimate the volume magnetization $M$, we used the bulk density of the sample (3.15 g/cm$^3$). The magnetic critical current density $J_c(H)$ in A/cm$^2$ was determined using the Bean critical state model for a rectangular sample:

$$J_c(H) = 20\Delta M / \left(a - \frac{a^2}{3b}\right), \quad (S2)$$

where $a$ and $b$ ($a \leq b$ in cm) are cross-sectional sample-dimensions perpendicular to the magnetic field, and $\Delta M$ is the difference between the upper and lower branches of the hysteresis loops in emu/cm$^3$.

**H. Estimation of the coherence length ξ and penetration depth λ.**

In type II superconductors, the upper critical field $H_{c2}$ is the field value for which the distance between two fluxons approaches the order of coherence length ξ. According to the Ginzburg-Landau (GL) theory, the relationship between $H_{c2}$ and ξ is given by [24]

$$H_{c2} = \Phi_0/(2\pi\xi^2), \quad (S3)$$

where $\Phi_0$ is a flux quantum ($\Phi_0 = h/2e \sim 2.1 \times 10^{-15}$ Wb). If we apply this relationship to our bulk like superconducting system, we obtain a value of $\xi = 5.7$ nm for $H_{c2}(0)=97.9$ kOe, which is estimated from the temperature dependence of the $H_{c2}(T)$ data shown in Fig. S2. The GL theory also predicts the following relationship between $H_{c1}$ and $\kappa = \lambda/\xi$ for a high-κ superconductor ($\kappa \gg 1$) [24],

$$H_{c1} = \frac{\Phi_0}{4\pi\lambda^2}\ln\kappa. \quad (S4)$$

From the above equation, λ is calculated to be 77.7 nm for $\xi = 5.7$ nm and $H_{c1}(0) = 713$ Oe [Fig. S2]. This value of λ is consistent with the initial assumption of κ~16 ≫ 1. Thus, we consider that the high-κ approximation is justified in our case.

**I. Magneto-optical (MO) imaging.**

Our MO imaging system is based on the Faraday rotation of a polarized light beam illuminating an MO active indicator film that we place directly on top of the sample's surface. The details of the experimental set up were reported in our previous paper [25]. We used a garnet-based field-sensing layer as a Faraday active indicator. The MO signals were detected ~3 μm above the sample surface. The rotation angle increases with the magnitude of the local magnetic field perpendicular to the sample. We can directly visualize and quantify the field distribution across the sample area by using crossed polarizers in an optical microscope. The images were recorded with a sCMOS camera (pco.pnada4.2, PCO) and transferred to a computer for processing. The gray level of the image was converted into the magnetic field value based on a calibration of the response of the Faraday indicator to a range of controlled perpendicular magnetic field as monitored by the sCMOS camera through the microscope. An external out-of-plane magnetic field was applied using an electromagnet with a field range up to 400 Oe. The temperature of the sample was controlled using a He-flow cryostat with ±0.1 K stability.

**J. Local current distribution obtained by an inversion of Biot-Savart law.**
The calculation of current components from the measured perpendicular component of the magnetic field $B_z$ is possible by using an inversion of Biot-Savart law. Several attempts to solve this problem for 1-dimensional (1D), 2D and 3D samples were reported [26,27]. As for the problem in 2D, a general solution was developed by Jooss *et al.* [28] and has been recently refined by Zuber *et al.* [29], as given by the following expression:

$$\tilde{J}_x(k_x, k_y) = \frac{ik_y}{\mu_0} e^{kh} \text{cosech}\left(\frac{kd}{2}\right) \tilde{B}_z(k_x, k_y, h, d), \quad (S5)$$

$$\tilde{J}_y(k_x, k_y) = \frac{-ik_x}{\mu_0} e^{kh} \text{cosech}\left(\frac{kd}{2}\right) \tilde{B}_z(k_x, k_y, h, d), \quad (S6)$$

where $k_x$ and $k_y$ are Fourier space wave-vectors, $k = \sqrt{k_x^2 + k_y^2}$, $\tilde{B}_z$ and $\tilde{J}_{x,y}$ are magnetic field and sheet current in Fourier space, respectively, $h$ is the distance between the top of the sample and the bottom of the indicator film, $d$ is the thickness of the sample, and $\mu_0$ is the permeability of free space. Accordingly, two current components $J_x$ and $J_y$ can be obtained by inverse Fourier transformation of equations (S5) and (S6). Here, it should be noted that equations (S5) and (S6) hold exactly for very thin samples, i.e., $h \gg d$, in which the perpendicular component of the current density $J_z$ does not contribute to $B_z$. However, even for thick samples, the measured $B_z$ is mainly sensitive to the surface currents within a surface layer of thickness of ~10 μm [30]. This indicates that MO and SSM imaging methods act like a microscope with finite depth of focus [30,31]. Thus, the 2D Fourier transformation based on the inverted Biot-Savart law can also be used to calculate, although semi-quantitatively, the current distribution in a near-surface layer of thick samples. In this work, we used the values of 3 and 6 μm for $h$ and $d$, respectively, in equations (S5) and (S6) to obtain $J_x$ and $J_y$. We confirmed that the pattern of the current distribution is not sensitive to the particular choice of $h$ and $d$ for as long as $h < 8$ μm and $d < 10$ μm. We therefore consider that the thus obtained current distribution pattern will represent the underlying nature of the vortex current flow in the near-surface region of the sample.

**K. AC magnetization measurements.**
The ac magnetization measurements were carried out for a square cuboid shape sample with dimensions of 1×1×5 mm$^3$ using a commercial SQUID magnetometer (MPMS 3, Quantum Design). Temperature variations of ac susceptibility were monitored in fixed temperature mode with the FC protocol under different amplitudes, and frequencies of the ac driving field and a constant dc bias magnetic field. When measuring the ac susceptibilities at different frequencies, the amplitude of the ac driving field was set to 1 Oe.

**L. Cathodoluminescence (CL) measurements.**

CL spectra and images were recorded using a cathodoluminescence spectrometer system (MP32 CL, Horiba) attached to a field emission scanning electron microscope (SU6600, Hitachi). The acceleration voltage was set to 5 or 10 kV, and the probe current was 3.6 or 7.2 nA according to the measurement requirements. The CL grating is 100 grooves/mm in order to cover a wide wavelength range (300–900 nm), and all measurements were conducted at room temperature.

**Supplemental Figures**

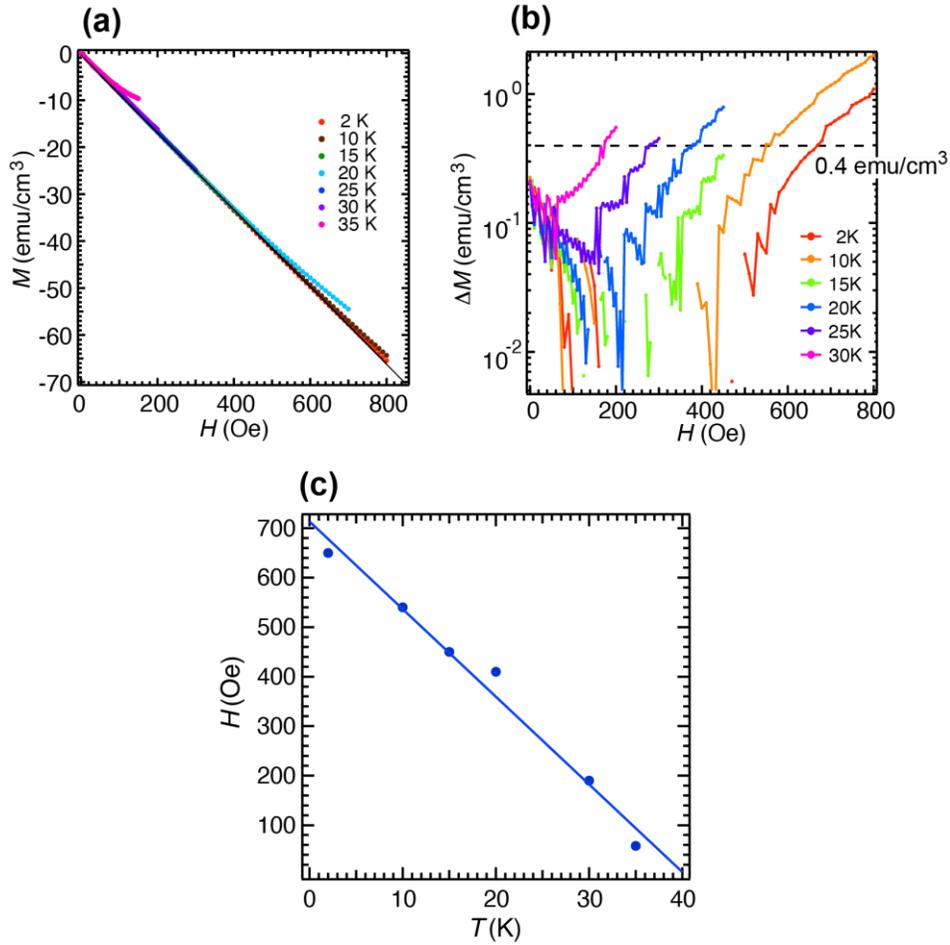

FIG. S1. (a) Initial zero-field cooling $M(H)$ curves obtained at different temperature. The solid line is the linear Meissner line. The value of $H_{c1}$ can be determined by examining the point of departure from the Meissner line on the initial slope of the $M(H)$ curve. (b) The difference between the linear line and the $M(H)$ curves. $\Delta M$ is shown in the logarithmic scale. $H_{c1}$ was acquired using the criteria of $\Delta M = 0.4$ emu/cm$^3$. (c) The temperature dependence of the $H_{c1}$. The solid line is a linear fit to the data. The extrapolated value at zero temperature $H_{c1}(0)$ is estimated to be 713 Oe.

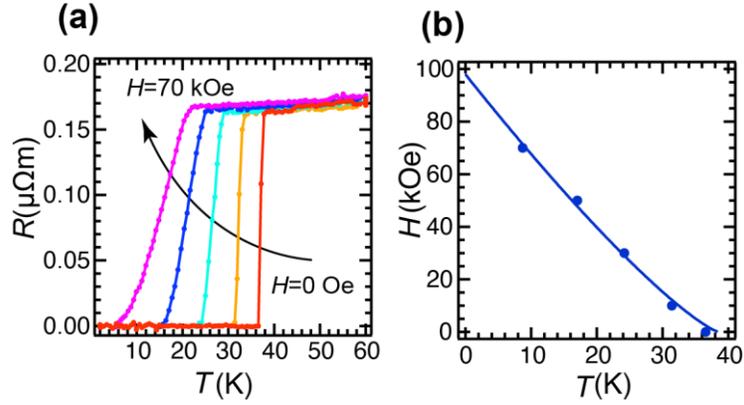

FIG. S2. (a) Resistive superconducting transition in different applied fields (from bottom right to top left) at 0, 10, 30, 50, and 70 kOe. Hc2(T) was defined as the applied field for which the sample resistance measured at T is 10% of the normal state value. (b) The temperature dependence of Hc2. The solid line represents fit of function $H_{c2}(T) = H_{c2}(0)(1 - T/T_c)^{1+\alpha}$. The fitted values of $H_{c2}(0)$ and $\alpha$ are 97.9 kOe and 0.23, respectively.

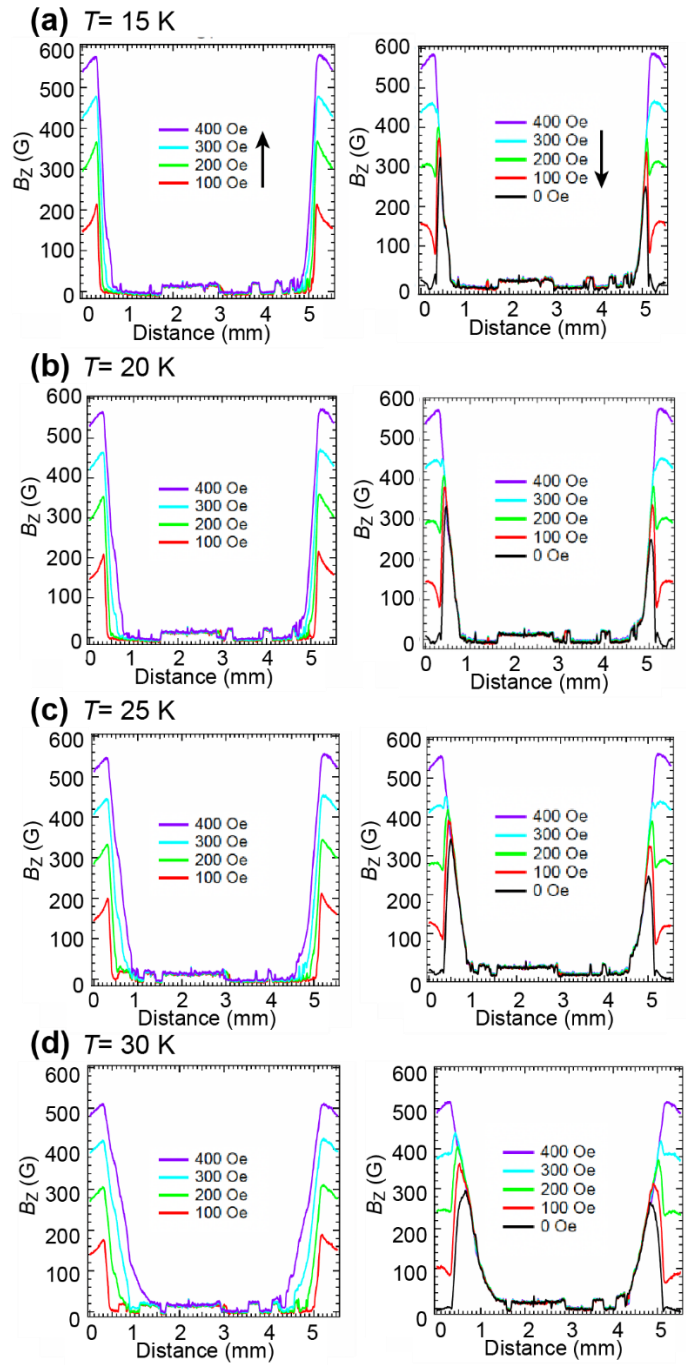

FIG. S3. *(continued to the next page)*

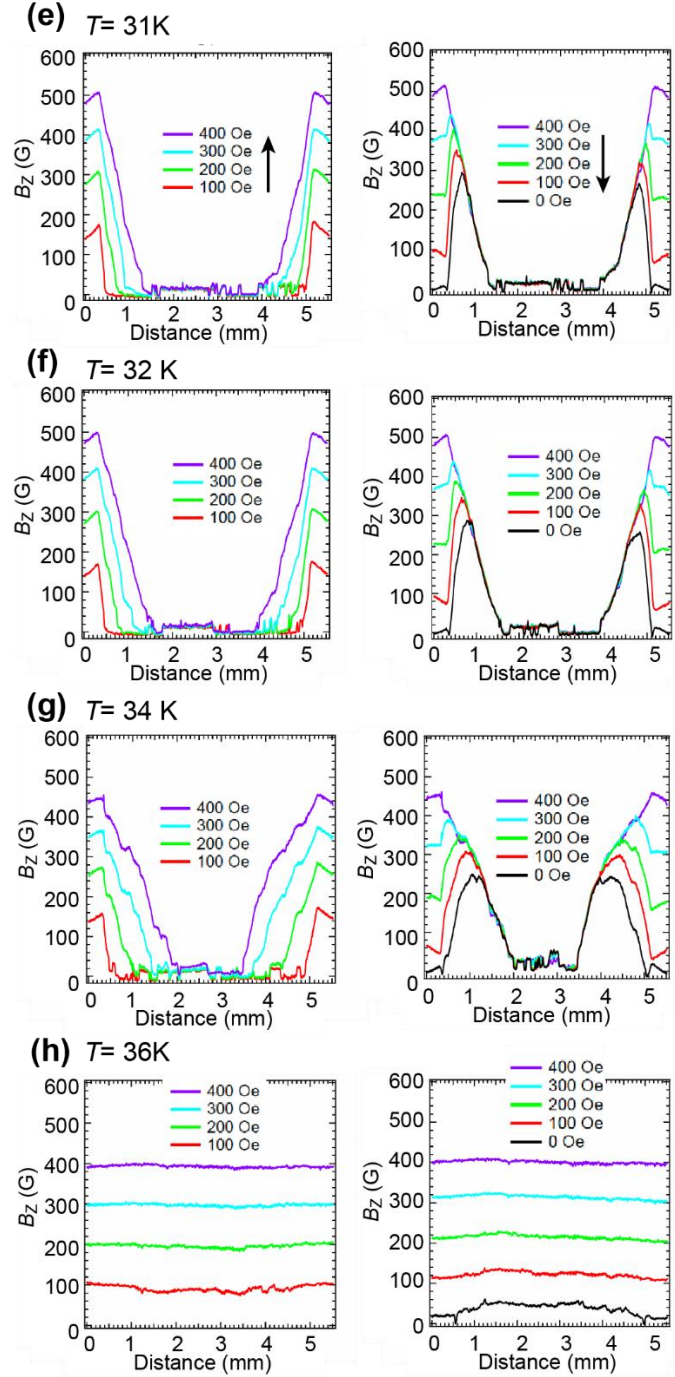

FIG. S3. (a)-(h) Profiles of perpendicular magnetic flux density $B_z$ obtained from the MO images acquired at different temperatures during the ramp-up stage of $H$ from 100 to 400 Oe (left panels) and the ramp-down stage of $H$ from 400 to 0 Oe (right panels).

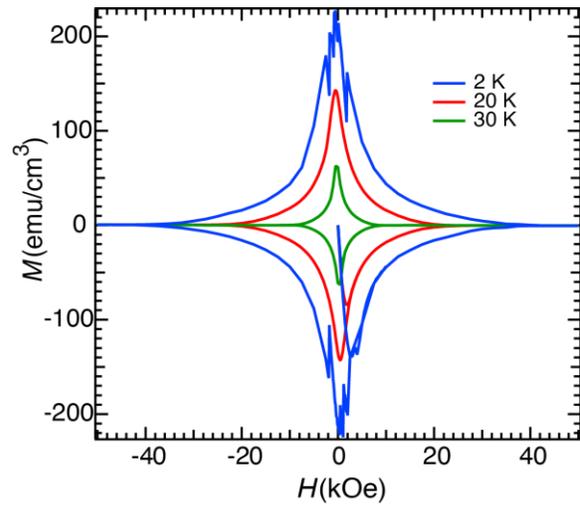
FIG. S4. *M*(*H*) hysteresis loops obtained at different temperatures.